# Large-signal model of the bilayer graphene field-effect transistor targeting radio-frequency applications: theory versus experiment

*Francisco Pasadas*[*] *and David Jiménez*

Departament d'Enginyeria Electrònica, Escola d'Enginyeria, Universitat Autònoma de Barcelona, 08193-Bellaterra, Spain

**ABSTRACT**

Bilayer graphene is a promising material for radio-frequency transistors because its energy gap might result in a better current saturation than the monolayer graphene. Because the great deal of interest in this technology, especially for flexible radio-frequency applications, gaining control of it requires the formulation of appropriate models for the drain current, charge and capacitance. In this work we have developed them for a dual-gated bilayer graphene field-effect transistor. A drift-diffusion mechanism for the carrier transport has been considered coupled with an appropriate field-effect model taking into account the electronic properties of the bilayer graphene. Extrinsic resistances have been included considering the formation of a Schottky barrier at the metal-bilayer graphene interface. The proposed model has been benchmarked against experimental prototype transistors, discussing the main figures of merit targeting radio-frequency applications.

---

[*] Electronic mail: Francisco.Pasadas@uab.cat

## I. INTRODUCTION

Graphene based devices have attracted the attention of scientific community due to the multitude of properties and the prospect of ultrahigh carrier mobilities at room temperature exceeding those of the conventional semiconductors. This has motivated intensive work focused on the development of high-frequency graphene metal-oxide-semiconductor field-effect transistors[1–3].

The gapless nature of monolayer graphene (MLG), which is a main obstacle on its application in graphene based electronics, causes the gate voltage to lose its control on switching off the device and are not suited for logic applications. However, it is currently believed that radio-frequency (RF) is the main electronic application where graphene can play a relevant role, having shown the highest carrier mobility and saturation velocity of any field-effect transistor (FET) channel material so far. A benchmarking of RF transistors considering a number of technologies has been reported in [1]. Both cut-off and maximum oscillation frequencies are behind other more mature technologies, mainly due to issues with lack of current saturation, large values of the contact and access resistances, together with suppression of carrier mobility because of the poor interface between the graphene layer and gate high-κ dielectrics. To overcome those issues, a number of different ideas are being explored. For instance, (i) the use of organic polymer buffers between the graphene and gate dielectrics in order to preserve the carrier mobility of the graphene[4], (ii) the use of edge contacts to minimize the S/D contact resistance[5] and (iii) relying on T-gate architectures to reduce the access resistance and gate fringing capacitance[6]. State-of-the-art graphene-based field effect transistors (GFETs) can deliver cut-off frequencies up to 427 GHz[7] and maximum frequency of oscillation of 105 GHz[8]. That large cut-off frequency is comparable with High-Electron Mobility Transistors (HEMTs) based on InP and GaAs, but the maximum oscillation frequency is considered low and still lies well behind III-V and Si-based transistors[9]. In particular, the absence of a band gap in graphene may prevent proper current saturation, especially at the required short gate lengths. Thus, introducing it in graphene is necessary. An interesting possibility is to get the band gap through size quantization. That is feasible using graphene nanoribbons[10,11], for which gaps up to 2.3 eV [12] have been demonstrated. Another alternative is offered by the bilayer graphene (BLG), where a gap can be induced either by molecular doping[13,14] or by applying a potential difference between two layers as a result of an external perpendicular electric field. Moreover, that potential difference can be realized with an applied gate field, meaning that the gap can be controlled by gate bias[15–18].

In this context, it is relevant to have predictive models for quantitative exploration of performances. Specifically, for the monolayer graphene field-effect transistor (MLGFET), several large-signal models based on a drift-diffusion (DD) theory have been proposed[19–25]. Besides, in order to investigate the performance offered by bilayer graphene FETs (BLGFETs), several models have been developed so far, e.g. Ryzhii *et al*. presented an analytical one based on the Boltzmann kinetic equation and Poisson equation in the weak nonlocality approximation[26,27]; Cheli *et al.* proposed an analytical model based on the effective mass approximation to calculate the thermionic and interband tunneling components of the current under the ballistic transport assumption[28]; Ghobadi and Abdi investigated the device characteristics by calculating the transmission coefficient through a tight-binding method[29]; and Fiori and Iannaccone carried out a study of the main RF figures of merit (FoMs) of a BLGFET through the NanoTCAD ViDES simulator, based on the self-consistent solution of the three-dimensional Poisson and Schrödinger equations by means of the non-equilibrium Green's function formalism[30]. The ballistic assumption in which all these models rely on seems unrealistic for the prototype devices explored so far, which do not fulfill the condition $L \ll \lambda$, where $L$ refers to the transistor channel length and $\lambda$ is the so called mean-free-path. The latter has been estimated as $\lambda \approx 10$ nm at carrier densities of $3 \times 10^{12}$ cm$^{-2}$ for exfoliated BLG deposited on a 300 nm $SiO_2$ substrate[31]. Hence it is worth reconsidering the carrier transport issue under the light of a DD theory when dealing with the practical situation $L \gg \lambda$. So in this work we present appropriate DD based models for the drain current, charge and capacitance of dual-gated BLGFETs, pursuing the following goals: (i) performance assessment and benchmarking against other existing technologies, (ii) provide guidance for device design, and (iii) bridge the gap between device and circuit level. The model starts by considering the device electrostatics. For such a purpose the two-dimensional density of states (2D-DOS) of BLG has been extracted from an effective two-band Hamiltonian at low energy. Upon application of Gauss's law to the gate stack, the carrier concentration and the electrostatic potential of the bilayer can be determined as a function of applied gate bias. Next, we have looked into the carrier transport considering a DD approach from which the drain current model can be formulated. Based upon it we have derived the charge associated to each transistor's terminal and a complete capacitive model, guaranteeing charge conservation, has been derived as a final step. Just to be sure the model captures the experimental evidence it has been validated against reported experimental results. Finally, main FoMs have been projected to illustrate the feasibility of using BLG for high frequency electronics.

## II. METHODS

In this section, we provide a detailed description of the large-signal model for the BLGFET. First of all, we present an investigation of the device's electrostatics, which forms the basis to later formulate the drain current, which is based on a DD approach. An important correction to the drain current that we have incorporated is the impact of the contact resistance after the formation of a Schottky barrier at the metal-BLG junction. Next, we have formulated appropriate models for the charge and capacitance, which are needed for any transient dynamics or frequency response simulation of the BLGFET. Finally we have assessed all these models against several prototype devices providing some quantitative predictions of the RF FoMs.

**A. Electrostatics.** BLG consists of two coupled monolayers of carbon atoms, each with a honeycomb crystal structure, arranged according to Bernal AB-stacked. The energy dispersion relation can be derived by means of an effective two-band Hamiltonian at low energy, that is, assuming that the intralayer hopping, $\gamma_0$, and the interlayer coupling, $\gamma_1$, are larger than other energies[32]; otherwise a four band model of the electronic bands is required to get the correct physical properties[33,34]. Based on the two-band Hamiltonian, the energy dispersion relation for BLG reads as:

$$E(k) = \frac{U_1+U_2}{2} \pm \sqrt{\frac{\gamma_1^2}{2} + \frac{U^2}{4} + v_F^2 k^2 \hbar^2 - \sqrt{\frac{\gamma_1^4}{4} + v_F^2 k^2 \hbar^2 (\gamma_1^2 + U^2)}} \quad (1)$$

where $U_1$ and $U_2$ are the potential energies of the first and second layers, respectively; $U = U_1-U_2$ is the interlayer asymmetry; $v_F = (3a\gamma_0/2\hbar)$, is the Fermi velocity; $a$ is the carbon-carbon distance of the BLG lattice structure; and $\hbar$ is the reduced Planck's constant. The above energy dispersion relation results in two bands with "Mexican hat" like shape and a gap of $E_{gap} = |U|\gamma_1/\sqrt{\gamma_1^2+U^2}$. From it, the band gap and the 2D-DOS at low energy can be derived:

$$DOS_{2D}(E) = \frac{2|E|}{\pi \hbar^2 v_F^2} \left(1 + \frac{1}{2}\sqrt{\frac{U^2+\gamma_1^2}{E^2-E_c^2}}\right) \quad (2)$$

From the derived 2D-DOS both the $n$ and $p$ (-type) carrier concentration can be easily calculated assuming a carrier distribution based on the Fermi-Dirac statistics[28] and considering $E_F = -(U_1+U_2)/2$ the Fermi energy and $E_c = E_{gap}/2$ ($E_v = -E_{gap}/2$) the conduction (valence) band edge.

Let us consider a BLGFET with top and back gates, with the cross-section depicted in FIG. 1a. Upon application of Gauss's law to the double-gate stack shown in FIG. 1b, we can get the carrier density and potentials on each layer from the external gate bias and impurities concentration:

$$\begin{aligned} C_t(V_{gs} - V_{gs0} - V_1) + C_o(V_2 - V_1) &= -\sigma_1 \\ C_b(V_{bs} - V_{bs0} - V_2) + C_o(V_1 - V_2) &= -\sigma_2 \end{aligned} \quad (3)$$

where $C_t = \varepsilon_0\varepsilon_t/(L_t-c_0/2)$ and $C_b = \varepsilon_0\varepsilon_b/(L_b-c_0/2)$ are the top and bottom oxide capacitances, respectively; $V_{gs}-V_{gs0}$ and $V_{bs}-V_{bs0}$ are the top and bottom gate voltage overdrive. These quantities comprise work-function differences between the gates and the graphene channel and possible additional charge due to impurities or doping; $V_1$ and $V_2$ are the electrostatic potentials at the first and second graphene layer; $C_o = \varepsilon_0\varepsilon_g/c_0$ is the graphene parallel plate capacitance, where $c_0$ is the interlayer spacing and $\varepsilon_g$ is an effective dielectric constant for the BLG to characterize charge screening[35]; and $\sigma_1$ and $\sigma_2$ are the charge densities at the first and second graphene layer, respectively.

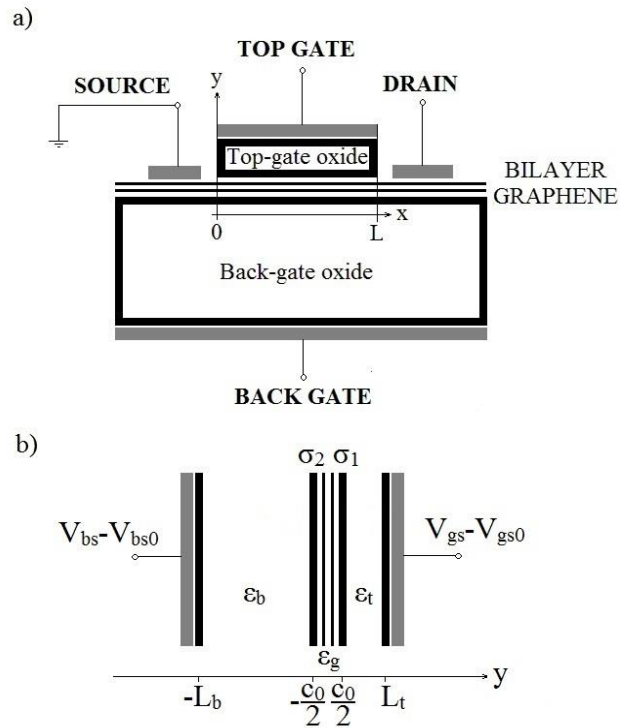

FIG. 1. a) Cross section of the BLGFET. It consists of two graphene sheets playing the role of the active channel. The electrostatic modulation of the carrier concentration in the 2D sheet is achieved via a double-gate stack consisting of top and bottom gate dielectrics and corresponding metal gates. The source is grounded and considered as the reference potential in the device. b) Scheme of the BLG based capacitor showing the relevant physical and electrical parameters, charges and potentials.

If the perpendicular electric field between the two graphene layers is assumed to be unscreened, then the charge density carried by each layer can be written as $\sigma_1 = \sigma_2 = Q_{net}/2$, where $Q_{net} = q(p-n)$ is the overall net mobile sheet charge density, where $q$ is the elementary charge. That simple assumption is known to overestimate the band gap[15,17]. A more accurate way of describing screening effects has been proposed by Edward McCann et al.[32] based on a tight-binding model and Hartree's theory. They have found that the individual layer densities are actually given by:

$$\sigma_{1(2)} = \frac{Q_{net}}{2} \mp \frac{q\gamma_1 U}{4\pi\hbar^2 v_F^2} \ln\left(\frac{\pi\hbar^2 v_F^2 |Q_{net}|}{2q\gamma_1^2} + \frac{1}{2}\sqrt{\left(\frac{\pi\hbar^2 v_F^2 Q_{net}}{q\gamma_1^2}\right)^2 + \left(\frac{U}{2\gamma_1}\right)^2}\right) \quad (4)$$

External gates are generally used to control the carrier density on a graphene device. For the BLG case, they also drive the separate layers to different potential energies $U_1$, $U_2$, inducing an interlayer asymmetry, $U$, and shifting of the Fermi energy, $E_F$. This physics can actually be explained in terms of displacement fields. A top and bottom electric displacement field, $D_t$ and $D_b$, respectively, built up upon application of top and bottom gate bias. The average of these quantities, $\Delta D = (D_b+D_t)/2$, breaks the inversion symmetry of the BLG and generates a nonzero band gap. The difference of both displacement fields, $\partial D = D_b-D_t$, shifts $E_F$ and creates a net carrier doping. At the point where $\partial D = 0$, named as charge neutral point (CNP), the Fermi level is located at the middle of the gap, and the corresponding electrical resistance is the highest. Those electric displacement fields (D) can be easily calculated as $D_b = \varepsilon_b(V_{bs}-V_{bs0})/(L_b-c_0/2)$ and $D_t = \varepsilon_t(V_{gs}-V_{gs0})/(L_t-c_0/2)$. In FIG. 2 we have illustrated how the applied gate voltages are tuning both the carrier density and the interlayer asymmetry and, ultimately, the band gap and the Fermi energy. The simulation was done using the parameters from Table I. The largest theoretical band gap that could be reached in BLG is limited by the intrinsic interlayer hopping parameter, $\gamma_1$. Experimentally, band gaps up to 250 meV have been reached[15].

The electrostatics of the BLGFET can be also represented using the equivalent capacitive circuit depicted in FIG. 3, which has been derived from Eq. 3 but replacing $V_{gs}$ and $V_{bs}$ by $V_{gs} - V(x)$ and $V_{bs} - V(x)$, respectively, where $V(x)$ is the quasi-Fermi level along the BLG channel. This quantity must fulfill the following boundary conditions: (1) $V(x) = 0$ at the source end, $x = 0$; (2) $V(x) = V_{ds}$ (drain-source voltage) at the drain end, $x = L$. The potential $V_c$ in the equivalent circuit represents the shift of the Fermi level (SFL) respect to the Dirac energy equivalent to the voltage drop across the quantum capacitance $C_q$, which is pretty the same concept that the surface potential in conventional silicon transistors. This quantity is usually defined as $C_q = dQ_{net}/dV_c$ and has to do with the 2D-DOS of the BLG. In nanoscale devices, where the oxide thicknesses could be small and the corresponding geometrical capacitances large, it could play a dominant role in defining the overall gate capacitance[36,37]. Both quantum capacitance and overall net mobile sheet charge of BLG have been presented in FIG. 4. Applying circuit laws to the equivalent capacitive circuit, the following straightforward relation is obtained:

$$V(x) = \frac{C_t}{C_t + C_b}\left(V_{gs} - V_{gs0} - V_1\right) + \frac{C_b}{C_t + C_b}\left(V_{bs} - V_{bs0} - V_2\right) + \frac{Q_{net}(V_c)}{C_t + C_b} \quad (5)$$

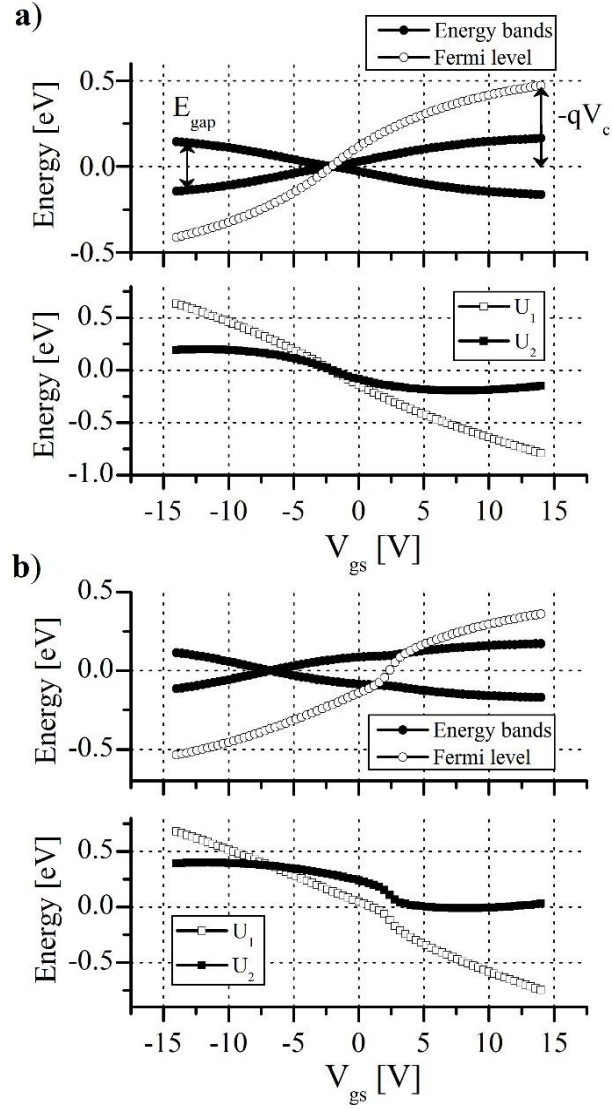

FIG. 2. Energy band diagrams and potential energies $U_1$, $U_2$ of a BLGFET as a function of the top gate bias for two different applied bottom gate bias: a) $V_{bs} - V_{bs0} = 0$ V; b) $V_{bs} - V_{bs0} = -40$ V. In the upper panel, the lines with black circles correspond to the conduction (upper) and valence (lower) band and the line with white circles represents the Fermi level. The voltage drop across the BLG, named as $V_c$, gives the position of the Fermi level respect to the CNP.

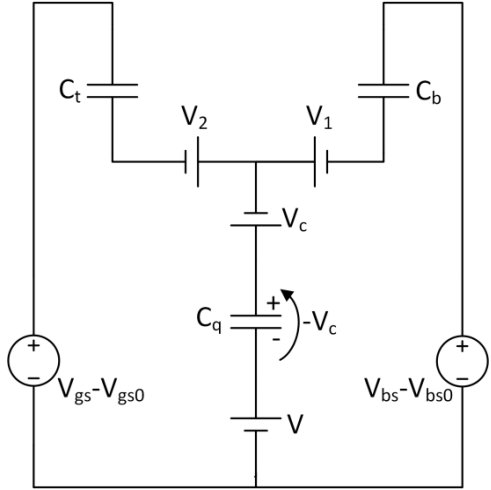

FIG. 3. Equivalent capacitive circuit of the BLGFET.

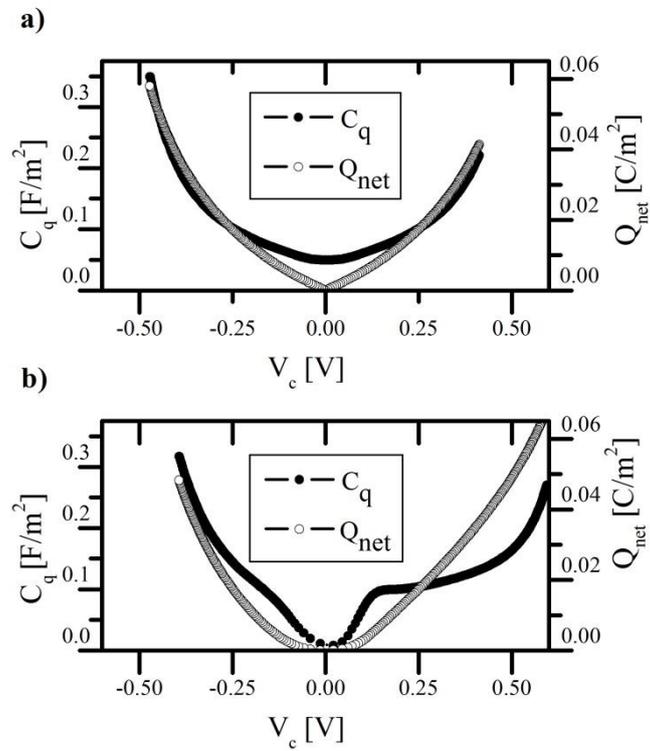

FIG. 4. Quantum capacitance and overall net mobile sheet charge density respect to the local electrostatic potential for two different applied bottom gate bias: a) $V_{bs} - V_{bs0} = 0V$; b) $V_{bs} - V_{bs0} = -40$ V. Theoretical results of the BLG quantum capacitance in case a) are consistent with calculations in [38,39].

**B. Drain current model.** To model the drain-to-source current of a BLGFET, a DD transport is assumed under the form $I_{ds} = -WQ_{tot}(x)v(x)$, where $W$ is the gate width, $Q_{tot}(x) = Q_t(x)+\sigma_{pud}$ is the free carrier sheet density along the channel at position $x$, $Q_t = q(p+n)$ is the transport sheet charge density, and $\sigma_{pud}$ is the residual charge density due to electron-hole puddles[31,40]. A soft saturation model was considered for the drift carrier velocity as $v(x) = \mu F(x)/(1+ \mu|F(x)|/v_{sat})$, where $F(x)$ is the electric field along the channel and $\mu$ represents the effective carrier mobility. We adopted this formulation consistently with the numerical studies of electronic transport in BLG relying on first-principles analysis and Monte Carlo simulations[41]. According to them, the saturation velocity can take a value between $0.25v_F$ and $v_F$. Applying $F(x) = -dV(x)/dx$, inserting the expression for $v(x)$ into the DD current equation and integrating over the device length, the drain current can be expressed as:

$$I_{ds} = \mu \frac{W}{L_{eff}} \int_0^{V_{ds}} Q_{tot} dV \qquad (6)$$

where $L_{eff} = L + \mu|V_{ds}|/v_{sat}$ is a correction to the physical channel length to incorporate saturation velocity effects. To get the drain current, it is convenient to solve the above integral using $V_c$ as the integration variable, and consistently express $Q_{tot}$ as a function of $V_c$ in the following way:

$$I_{ds} = \mu \frac{W}{L_{eff}} \int_{V_{cs}}^{V_{cd}} Q_{tot}(V_c) \frac{dV}{dV_c} dV_c \qquad (7)$$

where $V_{cs}$ and $V_{cd}$ are obtained from Eq. 5, with $V_{cs} = V_c|_{V=0}$ and $V_{cd} = V_c|_{V=V_{ds}}$. In addition, the quantity $dV/dV_c$ in Eq. 7 can also be derived from Eq. 5 and reads as follows:

$$\frac{dV}{dV_c} = -\frac{C_t}{C_t + C_b}\frac{dV_1}{dV_c} - \frac{C_b}{C_t + C_b}\frac{dV_2}{dV_c} + \frac{C_q}{C_t + C_b} \qquad (8)$$

To reproduce the experimental *I-V* characteristics of a BLGFET, accounting for the voltage drop at the source/drain (S/D) contacts is necessary. This quantity must be removed from external biases $V_{ds\_ext}$, $V_{gs\_ext}$ in order to get the internal ones $V_{ds}$, $V_{gs}$, respectively. State-of-the-art values for the metal-BLG contact resistance are around several hundred of $\Omega \cdot \mu m$[42–45]. A physical model of the contact resistance of metal-bilayer graphene junction has been described in Appendix A.

**C. Charge and capacitance models.** An accurate modeling of the intrinsic capacitances of FETs requires an analysis of the charge distribution in the channel versus the terminal bias voltages. So the terminal charges $Q_g$, $Q_b$, $Q_d$, and $Q_s$ associated with the top gate, bottom gate, drain, and source electrodes of a four-terminal device have been considered. For instance, $Q_g$ can be calculated by integrating $Q_{net\_g}(x) = C_t(V_{gs}-V_{gs0}-V_1(x)-V(x))$ along the channel and multiplying it by the channel width $W$. This expression for $Q_{net\_g}(x)$ has been obtained after applying Gauss's law to the top-gate stack as given by Eq. 9. A similar expression can be written for $Q_b$. It is worth noticing that $Q_g + Q_b = -W\int_0^L Q_{net}(x)dx$. On the other hand, the charge controlled by both the drain and source terminals can be computed based on Ward-Dutton's linear charge partition scheme, which guarantees charge conservation[19,46]. The resulting equations are listed next:

$$\begin{aligned}
Q_g &= WC_t\left[L(V_{gs}-V_{gs0}) - \int_0^L [V_1(x)+V(x)]dx\right] \\
Q_b &= WC_b\left[L(V_{bs}-V_{bs0}) - \int_0^L [V_2(x)+V(x)]dx\right] \\
Q_d &= W\int_0^L \frac{x}{L} Q_{net}(x)dx \\
Q_s &= -(Q_g + Q_b + Q_d)
\end{aligned} \quad (9)$$

The above expressions can conveniently be written using $V_c$ as the integration variable, as it was done to model the drain current. Based on the fact that the drain current is the same at any point $x$ in the channel (assuming there are no generation-recombination processes involved), we get from the DD transport model the following information:

$$x = \frac{\mu W}{I_{ds}} \left[ \int_{V_{cs}}^{V_c} Q_{tot}(V_c) \frac{dV}{dV_c} dV_c \right] \quad (10)$$

A four-terminal FET can be modeled with 4 self-capacitances and 12 intrinsic transcapacitances, which makes 16 capacitances in total. The capacitance matrix is formed by these capacitances where each element $C_{ij}$ describes the dependence of the charge at terminal $i$ with respect to a varying voltage applied to terminal $j$ assuming that the voltage at any other terminal remain constant.

$$C_{ij} = -\frac{\partial Q_i}{\partial V_j} \quad i \neq j$$
$$C_{ij} = \frac{\partial Q_i}{\partial V_j} \quad i = j \quad (11)$$

where $i$ and $j$ stand for $g$, $d$, $s$, and $b$.

$$\begin{bmatrix} C_{gg} & -C_{gd} & -C_{gs} & -C_{gb} \\ -C_{dg} & C_{dd} & -C_{ds} & -C_{db} \\ -C_{sg} & -C_{sd} & C_{ss} & -C_{sb} \\ -C_{bg} & -C_{bd} & -C_{bs} & C_{bb} \end{bmatrix} \quad (12)$$

Each row must sum to zero for the matrix to be reference-independent, and each column must sum to zero for the device description to be charge-conservative. Note that of the 16 intrinsic capacitances only 9 are independent.

## III. RESULTS AND DISCUSSION

The BLGFET drain-current, charge and capacitance models discussed so far will be assessed against different available experimental data. In the Experiment 1 we put under test the drain-current model of the BLGFET by comparing its prediction with the measured electrical behavior of a prototype device[16]. In the Experiment 2, all three models have been tested with the goal of highlighting the potential of using BLG to enhance RF performance metrics. The main RF FoMs used in this paper are described in Appendix B. So the experimental FoMs of the MLGFET reported in [47] have been compared with our theoretical prediction using an equivalent device that replaces the MLG by a BLG. Finally, in the Experiment 3, the outcome of our analytical models has been benchmarked against the electrical behavior of the BLGFET reported in [48], where an ON-OFF current ratio of 100 was measured.

**A. Experiment 1.** In this subsection we assess the drain-current model against the electrical characteristics reported in [16]. The simulations were done using the device's parameters listed in Table I. FIG. 5 shows both the experimental and predicted transfer characteristics (TCs) and output characteristics (OCs).

The mobility has been considered as an input parameter of the model to fit the experiment. It was assumed to be independent of the applied field, carrier density, or temperature, and considered the same for both electrons and holes. It is worth noting that some simulations and experiments have shown that the mobility somehow decreases with the size of the induced band gap[41,49], although we don't have included this refinement in our model.

The experimental TCs show a nonlinear shift of the CNP with the back-gate voltage. This effect is likely to appear because of the presence of charge traps in the gate oxide and/or the BLG interface. So when a positive $V_{bs}$-$V_{bs0}$ is applied to the device, the injection of electrons into the charge traps causes a shift of the CNP towards more positive voltage. On the contrary, applying a negative $V_{bs}$-$V_{bs0}$ results in hole injection, so the CNP shifts in the opposite direction. This effect has been reported in [50,51] for graphene on SiO$_2$ and the strength of it depends upon the swept voltage range, sweep rate, and surrounding conditions. So to capture this CNP shifting effect, a corrective parameter $\beta$ has been introduced in our model to properly modulate the top gate offset voltage, so $V_{gs0}$ was replaced by $V_{gs0}+\beta V_{bs}^2$, as proposed in [52]. Although the simulations do not match perfectly, the trends are actually well captured by the drain current model.

TABLE I. Input parameters of the models for experiment 1.

| Input parameter | Description | Value |
|---|---|---|
| $\gamma_0$[53] | In-plane hopping parameter | 3.16 eV |
| $\gamma_1$[53] | Interlayer hopping parameter | 0.381 eV |
| a[54] | Graphene lattice constant | 2.49 Å |
| $c_0$[54] | Graphene interlayer distance | 3.34 Å |
| $\beta$ | Fitting parameter due to nonlinearity in the response of the CNP to the back-gate bias | $3.2 \cdot 10^{-4}$ V$^{-1}$ |
| T | Temperature | 300 K |
| $\mu$ | BLG electron/hole mobility | 114 cm$^2$/Vs |
| L | Gate length | 4 μm |
| W | Gate width | 4 μm |
| $L_t$ | Top-gate oxide thickness | 8 nm |
| $L_b$ | Back-gate oxide thickness | 90 nm |
| $\varepsilon_g$[35] | Effective BLG relative permittivity | 2.5 |
| $\varepsilon_{top}$ | Top-gate oxide relative permittivity | 3.9 |
| $\varepsilon_{bottom}$ | Back-gate oxide relative permittivity | 3.9 |
| $V_{gs0}$ | Top-gate offset voltage | -2.1 V |
| $V_{bs0}$ | Back-gate offset voltage | -10 V |
| $L_c$ | Effective contact length | 4 μm |
| $A_{Ni-BLG}$ | Richardson constant for Nickel – BLG contact | $5 \cdot 10^4$ A/m$^2$K$^2$ |
| $v_{sat}$ | Saturation velocity | $0.75 \cdot v_F$ |

The electrostatics discussed in section II.A actually corresponds to the device under test. Specifically both FIG. 2a and FIG. 4a, depict a situation where the condition $D_b = 0$ (ZBDC, standing for Zero Bottom-gate electric Displacement field Condition) is fulfilled at $V_{bs} = V_{bs0} = -10V$, resulting in the brown curve shown in FIG. 5a. In this case, the charge neutrality condition is reached just at the zero-gap point, where the condition $D_t = 0$ is fulfilled at $V_{gs} = V_{gs0}$, so $\Delta D = \partial D = 0$. Increasing (reducing) the top gate bias beyond (below) the Dirac voltage results in the Fermi level directly entering into the conduction band, CB, (valence band, VB), so there is no any especial advantage of using BLG over MLG. On the other hand, when ZBDC does not apply, a larger current modulation can be obtained. The electrostatics of this situation is illustrated in FIG. 2b and FIG. 4b, where now $V_{bs}$-$V_{bs0}$ = -40V, which in turn corresponds to the green curve in FIG. 5a. In this case, the charge neutrality condition ($\partial D = 0$) is reached when $D_t = D_b \neq 0$, so $\Delta D \neq 0$ and the CNP happens at some finite energy band gap. Moving the top gate bias beyond (below) the CNP results in electron (hole) doping of the BLG together with an induced band gap that can reach a few hundred of meVs. Importantly, the Fermi level does not directly enter into the CB (VB) beyond (below) the CNP upon application of top gate bias, but there exists a region where it lies inside the band gap. So the combination of these two effects results in larger ON-OFF current ratio than the MLG based transistor.

The evolution of the Schottky barrier height seen by the electrons and holes at both S/D sides (see Appendix A) are shown in FIG. 5b as a function of $V_{gs\_ext}$. The corresponding $R_c$ is shown as well, broken down into its components $R_s$ and $R_d$, each calculated as the parallel association of the individual contact resistances due to electrons and holes. The bottom gate bias was fixed at $V_{bs}$ = -50V, far from the ZBDC. The corresponding electrostatics is plotted in FIG. 2b. It happens that the highest contact resistance situation is reached at $V_{gs\_ext}$ = -1.1 V and $V_{gs\_ext}$ = 0.8 V, corresponding to the pinch-off condition at the drain and source sides, respectively. The Schottky barrier height at these points is just $E_{gap}/2$, as given by Eq. A3.

Next, we show the experimental and simulated OCs of the BLGFET near to the ZBDC and far from it. As for the former situation, analysed in FIG. 5c left, saturation is weak, pretty similar to what is observed in MLG based transistors. On the contrary, biasing the device far from the ZBDC, results in current saturation over a sizeable range of $V_{sd\_ext}$ (FIG. 5c right). Simulations of $V_{cs}$, $V_{cd}$, CB's bottom, VB's top, all of them calculated as a function of $V_{sd\_ext}$, shown in FIG. 5d, are helpful to understand why that is happening. As for the near to ZBDC (FIG. 5d left) the pinch-off condition is reached when $V_{cd}$ = 0. This is happening at $V_{sd\_ext}$ = 1.2 V. Further increasing (reducing) of $V_{sd\_ext}$ drives the SFL at the drain side deep into the VB (CB), triggering the current due to holes (electrons). On the other hand, when the transistor is biased far from the ZBDC (FIG. 5d right), the pinch-off condition now occurs when the SFL at the drain side crosses the middle of a larger gap. This is happening at $V_{sd\_ext}$ = 1.4 V in the experiment. But now, for a moderate increase (decrease) of $V_{sd\_ext}$, the SFL at the drain side lies inside the gap, so there won't be appreciable current variation respect to the pinch-off condition, resulting in the observed current saturation. Eventually, if $V_{sd\_ext}$ is further increased (reduced) beyond (below) the range from 0.8 to 2V, then the SFL at the drain side enters into the VB (CB) and the device gets into the second (first) linear region dominated by holes (electrons). So the induced gap of the BLG provides a feasible way to virtually extend the pinch-off condition over a larger range of $V_{sd\_ext}$, which is of the utmost technological importance.

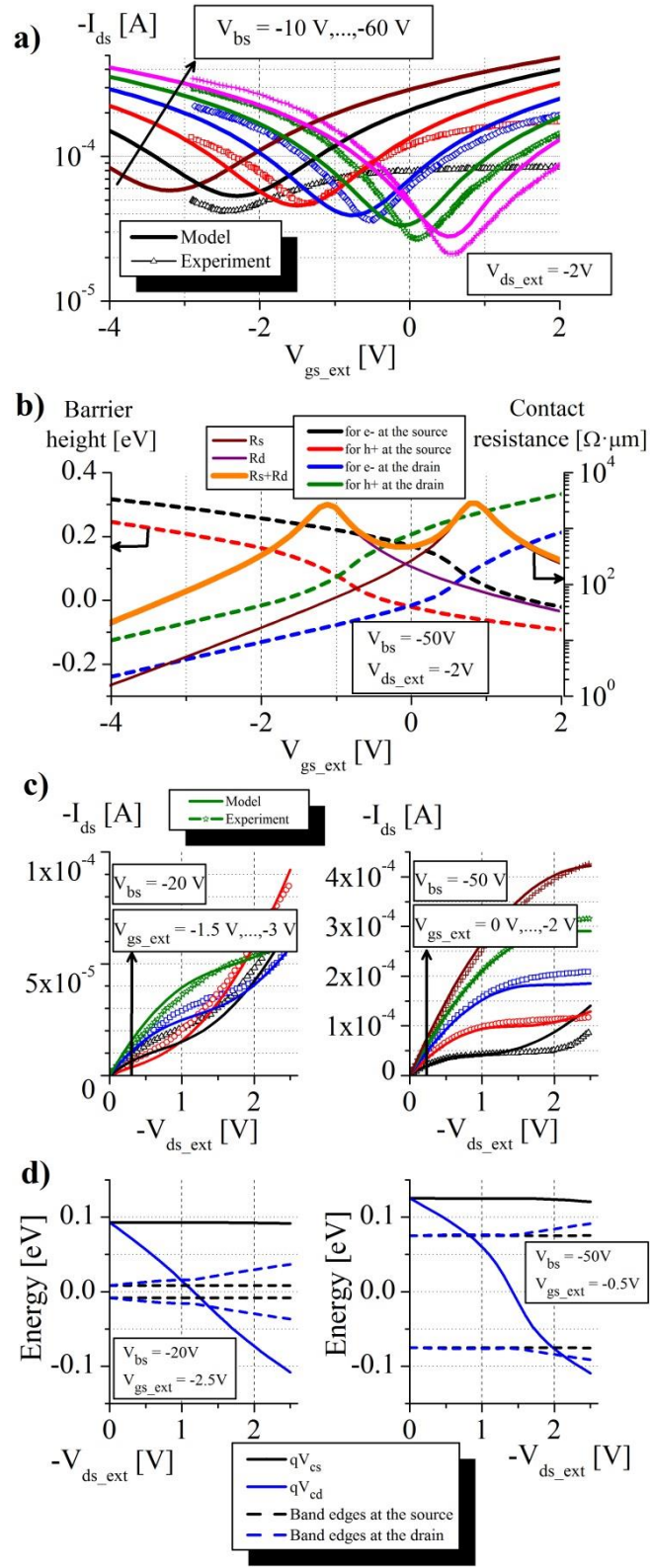

FIG. 5. (Color online) a) Transfer characteristics of the examined device. b) Schottky barrier height for both electrons and holes (left axis) and contact resistance (right axis) at the drain and source sides respect to the top gate bias. c) Output characteristics for two situations: (left) $V_{bs}$ = -20V; and (right) $V_{bs}$ = -50 V. d) Evolution of the SFL, conduction and valence band edges at the drain and source sides, with the drain bias according to the situations described in c).

**B. Experiment 2.** The goal of this subsection is to put under a new test the models presented so far and give some indications on the expected RF behavior of both MLG and BLG based transistors. As for the MLG case we rely on the experimental data of the GFET reported in [47] and the physics-based model developed in [19]. Both the experimental and simulated FoMs were compared with those derived from the BLGFET drain, charge and capacitance models. For such a comparison, the same input parameters have been used except those related with the material itself (see Table II).

TABLE II. Input parameters for experiment 2.

| Input parameter | Description | Value |
| --- | --- | --- |
| $\gamma_0$ | In-plane hopping parameter | 3.16 eV |
| $\gamma_1$ | Interlayer hopping parameter | 0.381 eV |
| a | Graphene lattice constant | 2.49 Å |
| $c_0$ | Graphene interlayer distance | 3.34 Å |
| $\beta$ | Fitting parameter due to nonlinearity in the response of the CNP to the back-gate bias | 0 V$^{-1}$ |
| T | Temperature | 300 K |
| $\mu$ | BLG electron/hole mobility | 400 cm$^2$/Vs |
| L | Gate length | 360 nm |
| W | Gate width | 40 μm |
| $L_t$ | Top-gate oxide thickness | 12 nm |
| $L_b$ | Back-gate oxide thickness | 300 nm |
| $\varepsilon_g$ | Effective BLG relative permittivity | 2.5 |
| $\varepsilon_{top}$ | Top-gate oxide relative permittivity | 7.5 |
| $\varepsilon_{bottom}$ | Back-gate oxide relative permittivity | 3.9 |
| $V_{gs0}$ | Top-gate offset voltage | -0.8 V |
| $V_{bs0}$ | Back-gate offset voltage | 0 V |
| $L_c$ | Effective contact length | 400 nm |
| $A_{Ti-BLG}$ | Richardson constant for Titanium – BLG contact | $8 \cdot 10^4 A/m^2 K^2$ |
| $v_{sat}$ | Saturation velocity | $0.75 \cdot v_F$ |

First of all, we have analyzed the intrinsic capacitances of the BLGFET as a function of both the gate and drain biases, shown in FIG. 6a. As for the $C - V_{gs\_ext}$ characteristics (FIG. 6a left), there are up to three singular points referred as A, B, C in FIG. 6b left. Say, for instance the self-capacitance $C_{gg}$, where all three points lie within the simulated $V_{gs\_ext}$ window. Other capacitances might require expanding the voltage range to get all of them. Point A is reached at $V_{g\_ext}$ such as $V_{cs} = 0$, so the pinch-off point is just at the source side and the channel is entirely p-type. Further increasing of $V_{gs\_ext}$ produces the shifting of the pinch-off point to the middle of the channel where now $V_{cd} = -V_{cs}$, so the half part of the channel close to the source becomes p-type and the other half part close to the drain n-type, resulting in point marked as B. If $V_{gs\_ext}$ is still further increased, the condition $V_{cd} = 0$ will eventually be reached at the point C. In this case, the pinch-off point has been shifted exactly at the source side and the channel is entirely n-type. Similar discussion could be made for the $C - V_{ds\_ext}$ characteristics shown in FIG. 6a right according to SFLs represented in FIG. 6b right. The behavior discussed so far regarding the intrinsic capacitances are qualitatively similar to that reported for the MLG case in [19], although quantitative details might differ.

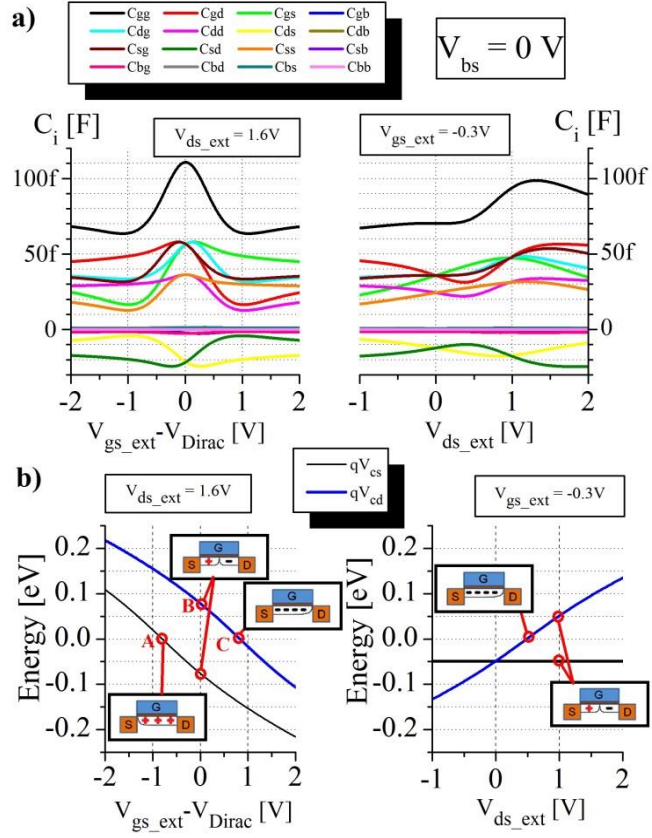

FIG. 6. (Color online) a) Intrinsic capacitances for the examined device versus the top gate bias (left) and drain bias (right), for $V_{bs}$ - $V_{bs0}$ = 0V. b) SFL at the drain and source sides plotted respect to top gate bias (left) and drain bias (right).

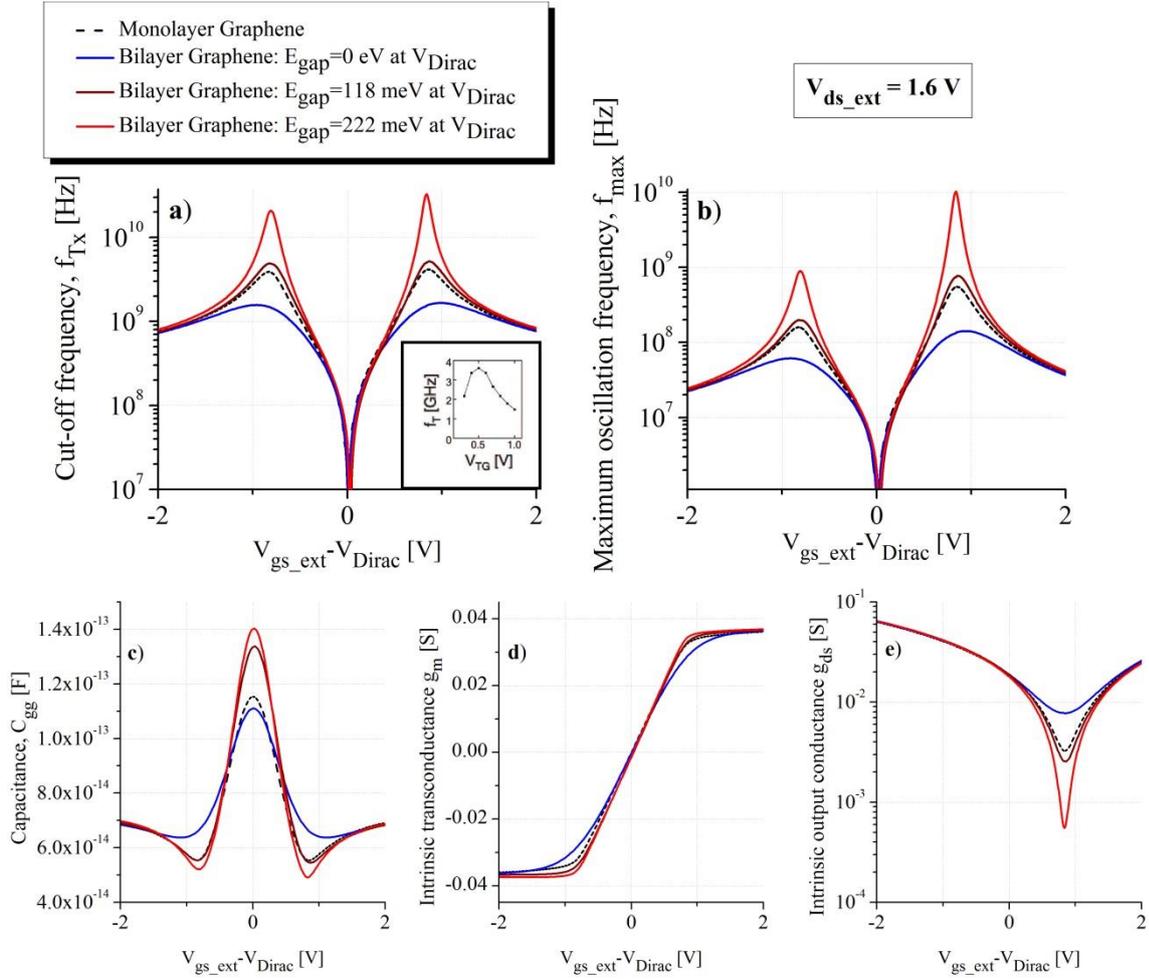

FIG. 7. (Color online) Theoretical calculation of the main RF figures of merit such as a) cut-off frequency $f_{Tx}$, and b) maximum oscillation frequency, $f_{max}$. These are shown for a MLGFET and BLGFET versus the top gate bias referenced to the Dirac voltage. Relevant parameters determining the FoMs behavior, such as c) the intrinsic $C_{gg}$ capacitance, d) intrinsic transconductance, $g_m$, and e) intrinsic output conductance $g_{ds}$ are also shown. The inset shows the experimental cut-off frequency measured for the MLGFET in [47].

The RF performance of graphene-based FETs has to do with the transconductance, output conductance, capacitances and extrinsic resistance as given by Eqs. B1 and B2 (see Appendix B). These equations emphasize the importance of minimizing all parasitic resistances, which have been taken into account in our contact resistance model explained before. A natural question arising is how far the BLG can go respect to its MLG counterpart regarding the RF performance. To answer this question we have considered a BLG channel with a variety of induced band gaps at the CNP. This can be done, in practice, by polarizing the device with appropriate $V_{bs}$. To start with, FIG. 7 shows the calculated $g_m$, $C_{gg}$ and corresponding $f_{Tx}$ as a function of $V_{gs\_ext}$. First observation is that $g_m$ looks like symmetric. This is because the equivalent role played by electrons and holes when positive and negative gate biases, respectively, are applied to the device. On the other hand, as the gate voltage is varied, $f_{Tx}$ is being modulated by $g_m$. Its value expands over several orders of magnitude depending on the top gate bias and reach up to several GHz in this example. The maximum takes place at $V_{gs\_ext}$ corresponding to the peak $g_m$, which is around 35 mS, in agreement with the experiment [47]. Moreover this bias point results in the minimum $C_{gg}$, so $f_{Tx}$ maximizes its value. Around this special point we can observe the clear advantage in using BLG instead of MLG, so when the induced gap is larger than 222 meV then $f_{Tx}$ scales up in a factor more than 8. Nevertheless, the slightly asymmetry in the two peaks of $f_{Tx}$ at negative and positive gate bias is due to the different output conductance.

Next, we look into the $f_{max}$ behavior in FIG. 7b. This FoM critically depends on how good the current saturation is and $g_{ds}$ is serving as key indicator. So to investigate it, we have also calculated $g_{ds}$ and $f_{max}$ vs $V_{gs}$ for different induced gaps, again by appropriate tuning of $V_{bs}$. As usual, the MLG case has been plotted as a reference. Provided that the gap is larger than one hundred meV, saturation becomes dramatically improved, so $f_{max}$ goes up to the maximum value (several GHz for the examined device). It is interesting to compare this result against the MLG case. The minimum $g_{ds}$ for the GFET, gotten at the CNP, is around 3.32 mS. However, for the BLGFET, when the gap size is 222 meV, becomes 6 times smaller and this value ultimately translates into 20 times larger $f_{max}$. This result highlights the importance of current saturation when it comes to optimizing RF FoMs.

**C. Experiment 3.** In this subsection we assess the drain-current model against the experimental TCs of the dual-gated BLGFET described in [48]. The TCs, shown in FIG. 8a, were recorded at room temperature by sweeping $V_{gs\_ext}$ while keeping constant $V_{bs}$. For comparison, the predicted TCs are shown in the same plot. The geometrical and electrical parameters used for the simulations are given in Table III. The model predicts a continuous enhancement of the ON-OFF current ratio expanding from 10 to 100 as $V_{bs}$ goes from 40 V down to -120 V, in correspondence to experimental evidence. According to our simulations the induced gap in the BLG at the CNP goes from 9.7 meV to 195 meV, and the maximum contact resistance goes from 150 Ω·μm to 2.6 kΩ·μm within the explored $V_{bs}$ range.

TABLE III. Input parameters of the models for experiment 3.

| Input parameter | Description | Value |
| --- | --- | --- |
| $\gamma_0$ | In-plane hopping parameter | 3.16 eV |
| $\gamma_1$ | Interlayer hopping parameter | 0.381 eV |
| a | Graphene lattice constant | 2.49 Å |
| $c_0$ | Graphene interlayer distance | 3.34 Å |
| $\beta$ | Fitting parameter due to nonlinearity in the response of the CNP to the back-gate bias | $1.28 \cdot 10^{-4}$ V$^{-1}$ |
| T | Temperature | 300 K |
| $\Delta$ | Spatial potential inhomogeneity due to electron-hole puddles | 0 eV |
| μ | BLG electron/hole mobility | 1160 cm²/Vs |
| L | Gate length | 3 μm |
| W | Gate width | 1.6 μm |
| $L_t$ | Top-gate oxide thickness | 19 nm |
| $L_b$ | Back-gate oxide thickness | 300 nm |
| $\varepsilon_g$ | Effective BLG relative permittivity | 2.5 |
| $\varepsilon_{top}$ | Top-gate oxide relative permittivity | 4.2 |
| $\varepsilon_{bottom}$ | Back-gate oxide relative permittivity | 3.9 |
| $V_{gs0}$ | Top-gate offset voltage | 0 V |
| $V_{bs0}$ | Back-gate offset voltage | 50 V |
| $L_c$ | Effective contact length | 3 μm |
| $A_{Ti-BLG}$ | Richardson constant for Titanium – BLG contact | $8 \cdot 10^4 A/m^2K^2$ |
| $v_{sat}$ | Saturation velocity | $0.75 \cdot v_F$ |

Regarding the OCs, no experimental data were reported in [48], so only the predicted OCs are shown in FIG. 8b. The left and right panel correspond to the OCs calculated at $V_{bs}$ = 40 V ≈ $V_{bs0}$ and $V_{bs}$ = -100 V << $V_{bs0}$, respectively. The induced band gap is 10 and 175 meV, respectively, so the minimum output conductance is reduced in a factor of 6.7 for the latter case. Finally, the predicted $f_T$ and $f_{max}$ are shown in FIG. 9 for both $V_{bs}$ under examination. Those are tunable with $V_{gs\_ext}$ showing a peak value of 3.7 and 5.2 GHz, respectively, with a noticeable improvement in a factor of 5 in $f_{max}$ and a factor of 2 in $f_{Tx}$ when the gap goes from 10 to 175 meV. Nevertheless, the device is not yet optimized and there is plenty of room to get higher FoMs. Scaling down of the channel length together with reducing the oxide thickness to keep short-channel effects under control is necessary.

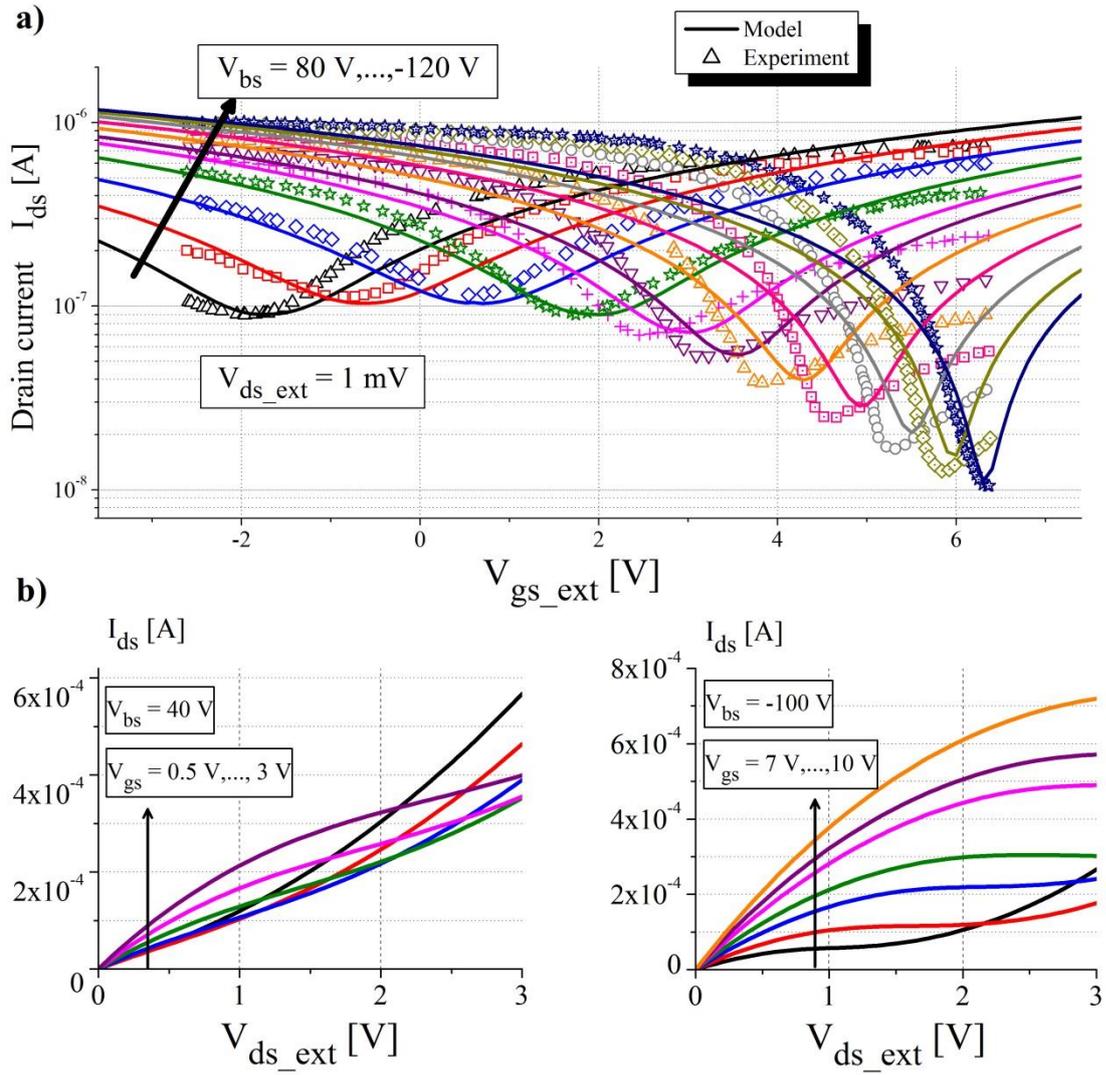

FIG. 8. (Color online) a) Transfer characteristics of the device under test. b) Output characteristics upon application of different $V_{bs}$ resulting in small and large band gap situations at the CNP: (left) $V_{bs}$ = 40V; and (right) $V_{bs}$ = -100 V.

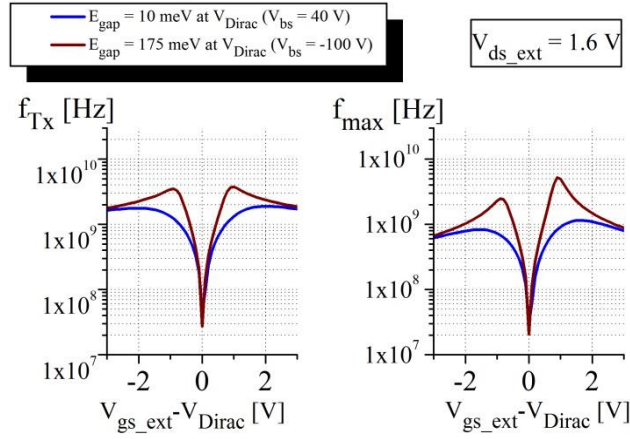

FIG. 9. (Color online) Prediction of a) cut-off frequency $f_{Tx}$, and b) maximum oscillation frequency, $f_{max}$, for the examined device upon application of different $V_{bs}$ resulting in small and large band gap at the CNP: (blue) $V_{bs}$ = 40V; and (brown) $V_{bs}$ = -100 V.

## IV. CONCLUSION

In summary, we have presented a large-signal model of the bilayer graphene field-effect transistor, including a thorough investigation of the drain current, charge and capacitances. It is based on a field-effect model and drift-diffusion carrier transport, including saturation velocity effects. The model makes full account of the tunable band gap nature of the bilayer graphene, which ultimately results into a better switch off the transistor together with enhanced drain current saturation as compared with the monolayer graphene counterpart. Extrinsic resistances have been included in the model considering the Schottky barrier formed between the metal contact and the bilayer graphene, which are known to degrade the RF performance. As for the capacitance model, we have used a Ward-Dutton's linear partition scheme that preserves charge conservation, which is critical for accurate transient simulations of circuits storing charge. The electrical behavior and RF performances of the BLGFET under diffusive transport regime, as far as we know, have not been modeled yet, although most of the prototype devices measured till now are likely to operate in this regime. The electrical models known so far relied on the ballistic assumption, exhibiting a lack of quantitative agreement. The large-signal model has been benchmarked against experimental prototype transistors, discussing the main figures of merit targeting radio-frequency applications. As for the considered bilayer graphene devices we have found enhancement factors up to 2 - 20 in either the cut-off frequency ($f_{Tx}$) or the maximum oscillation frequency ($f_{max}$) as compared with the equivalent monolayer graphene devices. These were gotten under application of an appropriate $V_{bs}$ bias producing a gap of some hundred meVs at the charge neutrality point. It is worth noting that the devices considered in this work were not optimized to get maximum performance. Optimization requires downscaling of the channel length together with appropriate choice of the insulator thickness and permittivity to keep short-channel effects under control. The scaling strategy to follow is unknown at the time being. Further investigation of this aspect is needed.

The model presented here is helpful to quantitatively estimate how the electronic properties of the bilayer graphene impact on the performance metrics of these devices. It is also a tool for the design of circuits for analog and radio-frequency applications. Inclusion of some additional physical effects as, for example, short-channel effects, nonquasi-static effects, trapped charges, carrier-density dependent mobility, might be necessary depending on the device geometry, fabrication process quality, and applied electrical excitation.

**ACKNOWLEDGMENTS**

We acknowledge Daniel Neumaier for providing us the data of Experiment 1. Support from Ministerio de Economía y Competitividad (MINECO) of Spain under project TEC2012-31330 and from the European Union Seventh Framework Programme under grant agreement nº 604391 Graphene Flagship are acknowledged.

**APPENDIX A: METAL-BILAYER GRAPHENE CONTACT RESISTANCE MODEL**

There are two physical processes responsible for the contact resistance: (1) carriers crossing the metal-BLG interface which hinders the carrier flow, namely $R_{mg}$, and (2) carriers flowing through an ungated BLG region until they reach the channel underneath the gate, $R_{gg}$. This latter component is sensitive to the gate voltage. A comprehensive model of the contact resistance has been reported by Chaves *et al.* [55,56]. The overall drain (source) parasitic resistance $R_D$ ($R_S$) is just $R_{mg} + R_{gg}$. In this work, we have only taken into account the contact resistance due to process (1), by assuming the formation of a Schottky barrier between the metal and the BLG. Inclusion of the gate-dependent contact resistance due to process (2) is feasible following the recipe given in [56]. Whenever an appreciable band gap exists, the current would be dominated by the thermionic emission of carriers through the Schottky barrier. Hence the current would be proportional to $exp(-q\phi_b/k_BT)$, where $\phi_b$ is the Schottky barrier height, $k_B$ is the Boltzmann constant, and $T$ is the temperature. So, the interfacial contact resistivity ($\rho_c$) between the metal and the BLG can be calculated as [57]:

$$\rho_c(\phi_b) = \frac{k_B}{qTA^*_{metal-BLG}} e^{\frac{q\phi_b}{k_BT}} \quad \text{(A1)}$$

where $A^*_{metal\text{-}BLG}$ is the Richardson constant of the metal–BLG contact, considered here as an empirical fitting parameter. The contact resistance can be expressed as [58]:

$$R_{c-Schottky}(\phi_b) = \frac{\sqrt{\rho_{sh}\rho_c(\phi_b)}}{W} \coth\left(L_c\sqrt{\frac{\rho_{sh}}{\rho_c(\phi_b)}}\right) \quad (A2)$$

where $L_c$ is the physical contact length and $\rho_{sh} = [q\mu(p+n)]^{-1}$ is the BLG sheet resistivity under the metal. It is possible to define a length $L_T = [\rho_c/\rho_{sh}]^{1/2}$ which physically corresponds to the length of the BLG region underneath the contact where the current mainly flows. Depending on the ratio between $L_c$ and $L_T$ two limit cases might arise: (1) short contact case ($L_c \ll L_T$), where the resistance is dominated by the interfacial contact resistance; and (2) long contact case ($L_c \gg L_T$), where the current flows uniformly across the entire contact. FIG. 10a shows the scheme of the physical structure of the metal-BLG contact together with an illustration of the current crowding phenomenon occurring for the short contact case.

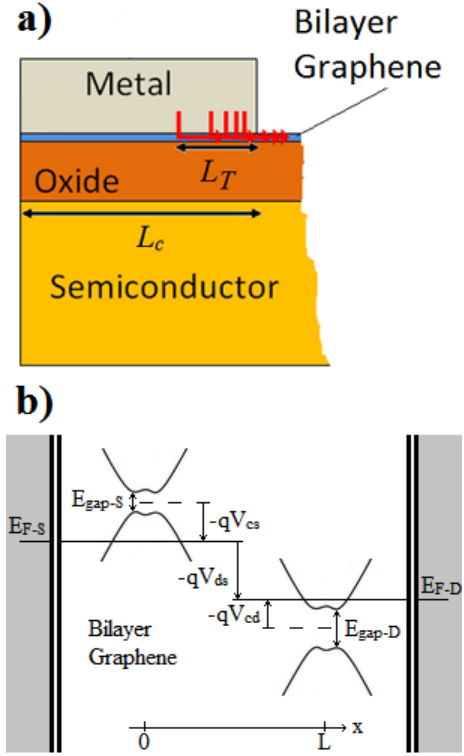

FIG. 10. (Color online) (a) Physical structure and scheme of the current crowding effect through the metal-BLG contact, (b) Schematics of the band diagram of the metal-BLG contact at the source and drain sides, needed to estimate the Schottky barrier height. The key quantities such as the band gap size, $E_{gap-S}$ and $E_{gap-D}$; the shift of the Fermi level, $V_{cs}$ and $V_{cd}$; and the Fermi energy at the metal, $E_{F-S}$ and $E_{F-D}$, both at the drain and source sides are shown. It is worth noticing that both $E_{F-S}$ and $E_{F-D}$ are aligned with the quasi-Fermi level at the source and drain sides, respectively; i.e. $E_{F-S} = V(0)$ and $E_{F-D} = V(L)$. The band diagram illustrates a possible mixed p/n-type channel with different band gap size on each side.

According to the band diagram shown in FIG. 10b the Schottky barrier height at the source side, which is presented separately for electrons, $\phi_b^n$, and for holes, $\phi_b^p$, can be calculated as:

$$\phi_{b-S}^n = V_{cs} + \frac{E_{gap-S}}{2}$$
$$\phi_{b-S}^p = -V_{cs} + \frac{E_{gap-S}}{2} \quad (A3)$$

An analogous procedure is implemented at the drain side. To quantitatively estimate the effect of the contact resistance, a splitting of the electron and hole contributions to the drain current is necessary. This can be done as follows:

$$I_{ds} = \mu \frac{W}{L_{eff}} \int_{V_{cs}}^{V_{cd}} Q_{tot}(V_c) \frac{dV}{dV_c} dV_c$$

$$= q\mu \frac{W}{L_{eff}} \left[ \int_{V_{cs}}^{V_{cd}} n(V_c) \frac{dV}{dV_c} dV_c + \int_{V_{cs}}^{V_{cd}} p(V_c) \frac{dV}{dV_c} dV_c \right] = I_{ds}^n + I_{ds}^p$$

where both $I_{ds}^n$ and $I_{ds}^p$ are the electron and hole contributions, respectively. The intrinsic $V_{gs}$ and $V_{ds}$ are then given by the following equations:

$$V_{gs} = V_{gs\_ext} - I_{ds}^n(V_{gs}, V_{ds}) R_{c-Schottky}(\phi_{b-S}^n)$$
$$- I_{ds}^p(V_{gs}, V_{ds}) R_{c-Schottky}(\phi_{b-S}^p)$$

$$V_{ds} = V_{ds\_ext} - I_{ds}^n(V_{gs}, V_{ds}) \left[ R_{c-Schottky}(\phi_{b-S}^n) + R_{c-Schottky}(\phi_{b-D}^n) \right]$$
$$- I_{ds}^p(V_{gs}, V_{ds}) \left[ R_{c-Schottky}(\phi_{b-S}^p) + R_{c-Schottky}(\phi_{b-D}^p) \right]$$

**APPENDIX B: RADIO-FREQUENCY FIGURES OF MERIT**

For RF transistors, the cut-off frequency ($f_{Tx}$) and the maximum oscillation frequency ($f_{max}$) are the most widely used FoMs to characterize the fundamental speed limit. The cut-off frequency is defined as the frequency for which the magnitude of the small-signal current gain of the transistor is reduced to unity. It is the highest possible frequency at which a FET is useful in RF applications. In general, the cut-off frequency $f_{Tx}$ is given by the following formula[59]:

$$f_{Tx} = \frac{|g_m|}{2\pi\left[(C_{gs} + C_{gd})(1 + g_{ds}(R_d + R_s)) + C_{gd}g_m(R_d + R_s)\right]} \quad (B1)$$

where $R_s$ and $R_d$ are the extrinsic resistances of the S/D contacts, respectively; $g_m$ ($=\partial I_{ds}/\partial V_{gs}$) is the intrinsic transconductance and $g_{ds}$ ($=\partial I_{ds}/\partial V_{ds}$) is the intrinsic output conductance. The cut-off frequency can be maximized by making the transconductance $g_m$ as large as possible and also making $g_{ds}$ as small as possible. It is also important for practical high frequency transistors, to make the device with the lowest possible extrinsic contact resistance.

On the other hand, the maximum oscillation frequency is defined as the highest possible frequency for which one can obtain power gain, namely, the frequency where the magnitude of the power gain of the transistor is reduced to unity. The $f_{max}$ can be written as[59]:

$$f_{max} = \frac{f_{Tx}}{2\sqrt{g_{ds}(R_g + R_s) + 2\pi f_{Tx}C_{gd}R_g}} \quad (B2)$$

where $R_g$ is the gate resistance. To get a high $f_{max}$, it is important to minimize all parasitic resistances in addition to having a high $f_{Tx}$. Additionally, the output conductance ($g_{ds}$) should be minimized, which is achieved through drain current saturation.